\documentclass[10pt,twocolumn,letterpaper]{article}
\usepackage[T1]{fontenc}
\usepackage[utf8]{inputenc}
\usepackage[letterpaper,margin=0.75in,columnsep=0.25in]{geometry}
\usepackage{microtype}
\usepackage{newtxtext,newtxmath}

\usepackage{tikz}
\usetikzlibrary{calc}
\usepackage{array}

\newcolumntype{C}[1]{>{\centering\arraybackslash}p{#1}}

\usepackage{amsmath,amsfonts}
\usepackage{algorithmic}
\usepackage{booktabs}
\usepackage{graphicx}
\usepackage{textcomp}
\usepackage{xcolor}
\usepackage[numbers,sort&compress]{natbib}
\usepackage{epsfig}

\usepackage{enumerate}

\usepackage{graphicx}
\usepackage{balance}
\usepackage{comment}
\usepackage{listings}
\usepackage{color,soul}
\usepackage{colortbl}
\usepackage{moreverb}
\usepackage{framed}
\usepackage{multirow}
\usepackage{pgfplots}
\usepackage{multirow}
\usepackage{rotating}
\usepackage{makecell}
\usepackage{pifont}
\usepackage{array}
\usepackage{subfig}

\usepackage{tcolorbox}
\usepackage{xurl}
\usepackage[hidelinks]{hyperref}
\usepackage{cleveref}

\usepackage{tikz}

\makeatletter
\newif\if@restonecol
\makeatother

\usepackage[ruled,vlined,linesnumbered]{algorithm2e}
\definecolor{lightgray}{gray}{0.9}
\definecolor{lightblue}{rgb}{0.9,0.9,1}
\definecolor{red}{rgb}{1,0,0}

\let\ls\lstinline

\newcolumntype{L}[1]{>{\raggedright\let\newline\\\arraybackslash\hspace{0pt}}m{#1}}

\usepackage{xspace}

\definecolor{bella@base03}{HTML}{002B36}
\definecolor{bella@base02}{HTML}{073642}
\definecolor{bella@base01}{HTML}{586e75}
\definecolor{bella@base00}{HTML}{657b83}
\definecolor{bella@base0}{HTML}{839496}
\definecolor{bella@base1}{HTML}{93a1a1}
\definecolor{bella@base2}{HTML}{EEE8D5}
\definecolor{bella@base3}{HTML}{FDF6E3}
\definecolor{bella@yellow}{HTML}{B58900}
\definecolor{bella@orange}{HTML}{CB4B16}
\definecolor{bella@red}{HTML}{DC322F}
\definecolor{bella@magenta}{HTML}{D33682}
\definecolor{bella@violet}{HTML}{6C71C4}
\definecolor{bella@blue}{HTML}{268BD2}
\definecolor{bella@cyan}{HTML}{2AA198}
\definecolor{bella@green}{HTML}{859900}
\author{
Chun Jie Chong, Muyeed Ahmed, Zhihao (Zephyr) Yao, Iulian Neamtiu \\
New Jersey Institute of Technology\\
{\normalsize\{cc255, ma234, zhihao.yao, ineamtiu\}@njit.edu}\\
}
\date{}
\AtBeginEnvironment{thebibliography}{\small\sloppy}

\makeatletter
\newcommand{\SetNode}[1]{\tikz[remember picture, overlay] \node (#1) {};}
\makeatother

\AtBeginDocument{}

\begin{document}

\title{Can LLMs be Effective Code Contributors? A Study on Open-source Projects}

\maketitle
\begingroup
\renewcommand\thefootnote{}
\footnotetext{Accepted to EASE 2026.}
\setcounter{footnote}{0}
\endgroup

\begin{abstract}
LLM-generated code is widely used, and the share of committed code produced by LLMs is expected to increase. However, we are not at a point where LLMs can be effective contributors to production code. 
We present an approach that exposes the shortcomings of LLM generation on such projects, and proposes recommendations; the targets of our study are sizable open-source projects, e.g., FFmpeg and wolfSSL.  
First, we developed a framework that uses verification and validation to evaluate a given LLM's suitability to fix or add features to an existing project.
Second, we apply the framework to 212 commits (bug fixes and small feature improvements) in eight popular open-source projects and three LLMs: GPT-4o, Ministral3, and Qwen3-Coder. The success rate varied from 0\% to 60\% depending on the project. The LLMs failed in a variety of ways, from generating syntactically incorrect code, to
producing code that fails basic (static) verification, 
or validation via the project's test suite. In particular, the LLMs struggle with generating new code, handling contexts (function or file) outside a certain size range, and in many cases their success is due to parroting code changes they have been trained on. 
\end{abstract}

\section{Introduction}
\label{sec:intro}

Thanks to Large Language Models (LLMs) advancements, AI-assisted programming is increasingly adopted in software development~\cite{github_study_ai_tool}.
AI code editors such as GitHub Copilot~\cite{github_copilot} (powered by 
LLMs including OpenAI's GPT~\cite{github_copilot_gpt4o}), Visual Studio IntelliCode~\cite{intellicode},
or Cursor~\cite{cursorai}, help programmers with code suggestions, explanations and implementation.
Accelerating software development is one of the reported benefits, e.g., a study found that 88\% of developers reported an increase in productivity when using Copilot~\cite{github_productivity}. 

However, 
LLM use comes with trade-offs. 
Studies have shown that LLMs misuse APIs~\cite{li2024assessing}, create irrelevant and complicated conditional logic~\cite{li2024assessing}, lack defensive programming~\cite{chong2024artificial}, 
introduce security issues~\cite{chong2024artificial}, 
and hallucinate~\cite{liu2024exploring}. Iterative prompting might fix some of these issues, but the results from iterative prompting are not consistent and not guaranteed~\cite{chong2024artificial, shukla2025security}. 
While benchmarks have been introduced to evaluate LLM-generated code~\cite{chen2021evaluating,liu2023repobench,yu2024codereval} they might not be representative of commits ``in the wild'' and LLMs could overfit to perform well on benchmarks alone.
These benchmarks also fail to capture the whole scope of real-world software evolution, e.g., code integration and testing (unit or whole-system).

We believe that an effective 
LLM must provide 
a fully automated solution that, {\it given a sizable code base and a prompt, generates a high-quality, usable commit} (patch) that satisfies the prompt specification and can be directly incorporated into the base. Towards that goal, several issues must be addressed:
(1) how to 
assess whether LLMs are suitable at generating commit-ready code, and 
(2) when LLMs fail, how do they fail, and what can we do about it.

To tackle these issues and help make LLMs more effective at contributing to sizable code bases, we developed an automated framework that evaluates LLMs' ability to fix bugs and implement new features via static analysis and test suites. Further comparisons between human-written and LLM-generated code allow us to identify LLM-generated code issues that escape the purview of automated verification\&validation.
Our automated framework uses programming tasks (commits) from open-source repositories, which are more representative than generic benchmarks. For our study, we focus on 212 actual commits (187 bug fixes and 25 feature enhancements) in 8 open-source projects: Bison, Collections-C, FFmpeg, jansson, libhl, packcc, Vsftpd, and wolfSSL. We focus on C code
as it is widely spread in system programming and open-source projects.

\Cref{sec:prelim} presents several motivational examples -- four actual tasks from commits -- along with the issues in the LLM-generated code.
\Cref{sec:approach} describes our approach: the framework, the projects and commits used for evaluation, and the verification\&validation processes.
\Cref{sec:findings} presents our study results along several Research Questions (RQs) designed to check LLMs' ability to generate effective code contributions, i.e., Can LLMs generate code that compiles, passes static analysis, and passes testing? When it fails, why and how does it fail? Which factors affect success rate?  

We applied our framework to three LLMs: GPT-4o, \mbox{Ministral3-14B}, and Qwen3-Coder-30B.
We now summarize some of our findings. Though the 212 commits were small, typically 4--15 LOC, the LLM-generated code fails to compile in 2--32 cases (depending on the LLM). Even when the code compiles, the verification stage found 9--18 null pointer dereferences, and 9--72 unsafe casts.
Among those 98--143 commits on which we could run a test suite, the LLMs had a passing rate of 71.8--86.7\%. The main causes for validation failure included partial fixes, empty generated patches, fixing the wrong thing, or deleting unrelated code.
We found that context size (subject file or target function) influences the success rate, e.g., a lower rate for larger contexts.
In \Cref{sec:discussions} we provide recommendations on how 
to increase effectiveness and safety when using LLM generation.

{\it Open Science policy.} Our artifact is available at\\
\url{https://github.com/LLM-Test-Framework/LLM-Test-Framework}.

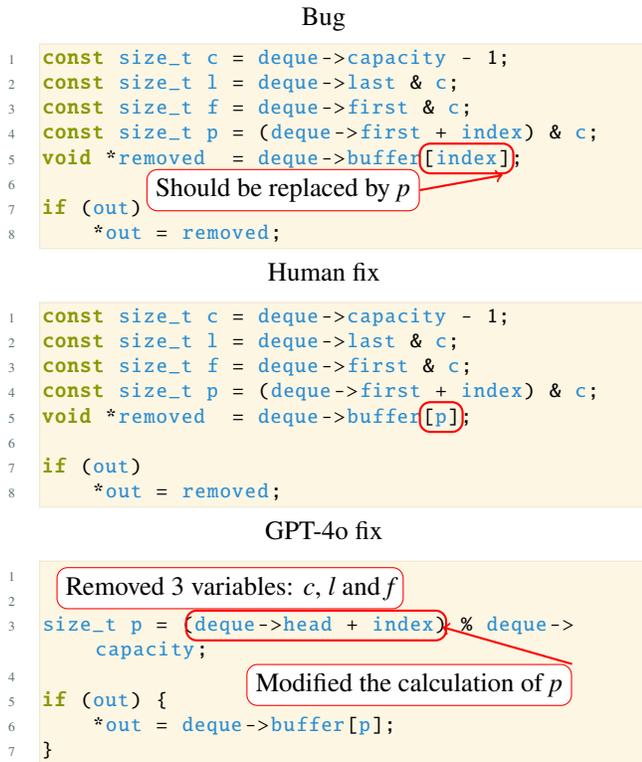
\begin{figure}[t]
\centering
\begin{minipage}{\columnwidth}
\centering
Bug
\begin{lstlisting}[xleftmargin=1.9em]
const size_t c = deque->capacity - 1;
const size_t l = deque->last & c;
const size_t f = deque->first & c;
const size_t p = (deque->first + index) & c;
void *removed  = deque->buffer[@\SetNode{a_start}@index@\SetNode{a_end}@]; 

if (out)
    *out = removed;
\end{lstlisting}
\begin{tikzpicture}[remember picture, overlay]
  \draw[red, thick, rounded corners]
    ($(a_start)+(-0.2,1.7ex)$) rectangle ($(a_end)+(0.2,-0.8ex)$);
  \node[draw=red, fill=white, rounded corners, anchor=east] 
    (label) at ($(a_start)+(-0.2,-0.35)$) {Should be replaced by {\it p}};
  \draw[->, thick, red] 
    (label.east) -- ($(a_start)+(0.9,-0.15)$);
\end{tikzpicture}
\end{minipage}\\[-4.5mm]
Human fix
\begin{minipage}{\columnwidth}
\vspace{3mm}
\begin{lstlisting}[{xleftmargin=1.9em}]
const size_t c = deque->capacity - 1;
const size_t l = deque->last & c;
const size_t f = deque->first & c;
const size_t p = (deque->first + index) & c;
void *removed  = deque->buffer[@\SetNode{b_start}@p@\SetNode{b_end}@];

if (out)
    *out = removed;

\end{lstlisting}
\begin{tikzpicture}[remember picture, overlay]
  \draw[red, thick, rounded corners]
    ($(b_start)+(-0.2,1.7ex)$) rectangle ($(b_end)+(0.2,-0.8ex)$);
\end{tikzpicture}
\end{minipage}\\[-4.5mm]
GPT-4o fix
\begin{minipage}{\columnwidth}
\vspace{3mm}
\begin{lstlisting}[{xleftmargin=1.9em}]
@\SetNode{c_1}@

size_t p = (@\SetNode{c2_start}@deque->head + index@\SetNode{c2_end}@) % deque->capacity;

if (out) {
    *out = deque->buffer[p];
}
\end{lstlisting}
\begin{tikzpicture}[remember picture, overlay]
  \node[draw=red, fill=white, rounded corners, anchor=west] 
    (label1) at ($(c_1)+(0.2,-0.1)$) {Removed 3 variables: {\it c}, {\it l} and {\it f}};
  \draw[red, thick, rounded corners]
    ($(c2_start)+(-0.1,1.7ex)$) rectangle ($(c2_end)+(0.15,-0.8ex)$);
  \node[draw=red, fill=white, rounded corners, anchor=west] 
    (label2) at ($(c2_start)+(0.7,-0.72)$) {Modified the calculation of {\it p}};
  \draw[->, thick, red] 
    (label2.north east) -- ($(c2_end)+(0.1,0.05)$);
\end{tikzpicture}
\end{minipage}
\caption{Human vs. GPT-4o fixes for an incorrect index bug.}
\label{fig:commit1}
\end{figure}

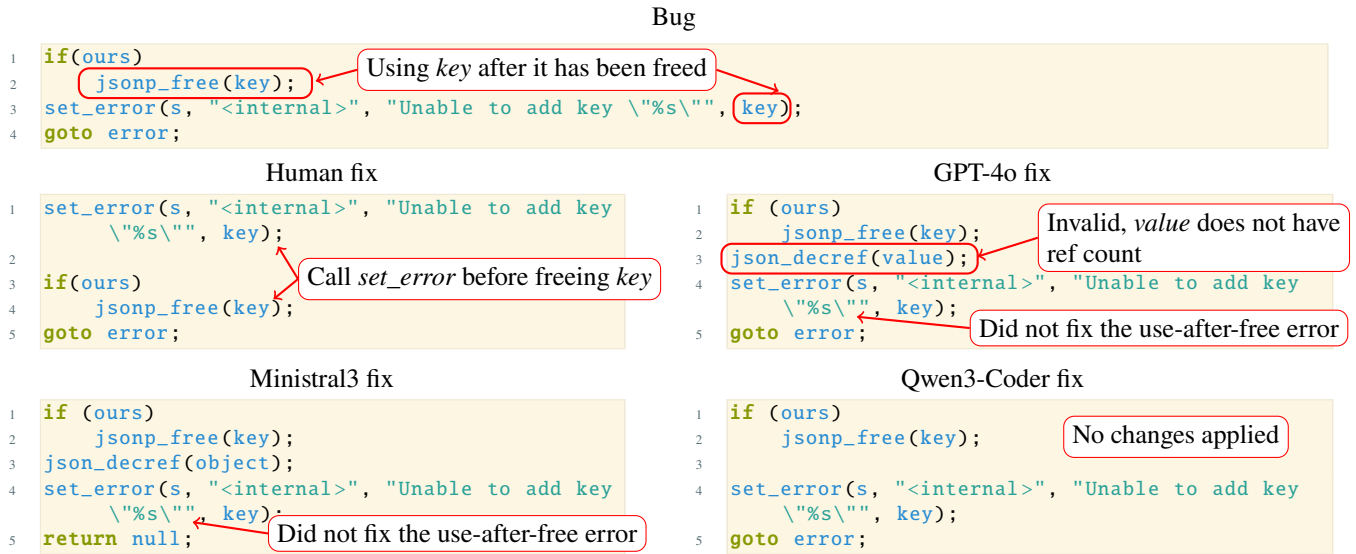
\begin{figure*}[t]
\centering
\begin{minipage}{0.975\textwidth}
\centering
Bug
\begin{lstlisting}[xleftmargin=1em]
if(ours)
    @\SetNode{a1_start}@jsonp_free(key);@\SetNode{a1_end}@
set_error(s, "<internal>", "Unable to add key \"%s\"", @\SetNode{a2_start}@key@\SetNode{a2_end}@);
goto error;
\end{lstlisting}
\begin{tikzpicture}[remember picture, overlay]
  \draw[red, thick, rounded corners]
    ($(a1_start)+(-0.2,1.7ex)$) rectangle ($(a1_end)+(0.2,-0.8ex)$);
  \draw[red, thick, rounded corners]
    ($(a2_start)+(-0.1,1.7ex)$) rectangle ($(a2_end)+(0.15,-0.8ex)$);
  \node[draw=red, fill=white, rounded corners, anchor=west] 
    (label) at ($(a1_start)+(3.5,0.25)$) {Using {\it key} after it has been freed};
  \draw[->, thick, red] 
    (label.west) -- ($(a1_end)+(0.25,0.1)$);
  \draw[->, thick, red] 
    (label.east) -- ($(a2_start)+(0.5,0.3)$);
\end{tikzpicture}
\end{minipage}\\[-5mm]
\begin{tabular}{cc}
Human fix & GPT-4o fix \\
\begin{minipage}{.45\textwidth}
\vspace{1mm}
\begin{lstlisting}[{xleftmargin=1em}]
set_error(s, "<internal>", "Unable to add key \"%s\"", key);@\SetNode{b1}@ 
@\SetNode{b}@
if(ours)
    jsonp_free(key);@\SetNode{b2}@
goto error;
\end{lstlisting}
\begin{tikzpicture}[remember picture, overlay]
  \node[draw=red, fill=white, rounded corners, anchor=west] 
    (label) at ($(b)+(3.4,-0.2)$) {Call {\it set\_error} before freeing {\it key}};
  \draw[->, thick, red] 
    (label.west) -- ($(b1)+(-0.1,-0.1)$);
  \draw[->, thick, red] 
    (label.west) -- ($(b2)+(-0.3,0.2)$);
\end{tikzpicture}
\end{minipage}
&
\hspace*{9mm}
\begin{minipage}{.45\textwidth}
\vspace{1mm}
\begin{lstlisting}
if (ours)
    jsonp_free(key); 
@\SetNode{c1_start}@json_decref(value);@\SetNode{c1_end}@
set_error(s, "<internal>", "Unable to add key \"%s\"", key);
@\SetNode{c2}@goto error;

\end{lstlisting}
\begin{tikzpicture}[remember picture, overlay]
  \draw[red, thick, rounded corners]
    ($(c1_start)+(-0.1,1.7ex)$) rectangle ($(c1_end)+(0.1,-0.8ex)$);
  \node[draw=red, fill=white, rounded corners, anchor=west, align=left] 
    (label) at ($(c1_end)+(0.9,0.35)$) {Invalid, {\it value} does not have\\ref count};
  \draw[->, thick, red] 
    (label.west) -- ($(c1_end)+(0.1,0.1)$);
  \node[draw=red, fill=white, rounded corners, anchor=west, align=left] 
    (label) at ($(c2)+(3.2,0.15)$) {Did not fix the use-after-free error};
  \draw[->, thick, red] 
    (label.west) -- ($(c2)+(1.7,0.3)$);
\end{tikzpicture}
\end{minipage} \\[-4mm]
Ministral3 fix & Qwen3-Coder fix \\
\begin{minipage}{.45\textwidth}
\vspace{1mm}
\begin{lstlisting}[{xleftmargin=1em}]
if (ours)
    jsonp_free(key); 
json_decref(object);
set_error(s, "<internal>", "Unable to add key \"%s\"", key);
@\SetNode{c2}@return null;

\end{lstlisting}
\begin{tikzpicture}[remember picture, overlay]
  \node[draw=red, fill=white, rounded corners, anchor=west, align=left] 
    (label) at ($(c2)+(3,0.15)$) {Did not fix the use-after-free error};
  \draw[->, thick, red] 
    (label.west) -- ($(c2)+(2.0,0.3)$);
\end{tikzpicture}
\end{minipage}
&
\hspace*{9mm}
\begin{minipage}{.45\textwidth}
\vspace{1mm}
\begin{lstlisting}
if (ours)
    jsonp_free(key); @\SetNode{c1_end}@
    
set_error(s, "<internal>", "Unable to add key \"%s\"", key);
goto error;

\end{lstlisting}
\begin{tikzpicture}[remember picture, overlay]
  \node[draw=red, fill=white, rounded corners, anchor=west, align=left] 
    (label) at ($(c1_end)+(0.9,0.1)$) {No changes applied};
\end{tikzpicture}
\end{minipage}
\end{tabular}
\caption{LLM-generated code for fixing a use-after-free bug.}
\label{fig:commit2}
\end{figure*}

\begin{figure*}[t]
\centering
\begin{minipage}{0.975\textwidth}
\centering
Bug
\begin{lstlisting}[xleftmargin=1em]
while (replace) {
    if (table->key_cmp(@\SetNode{a_start}@replace->key@\SetNode{a_end}@, key) == 0) {
        replace->value = val;
        return CC_OK;
    }
}
\end{lstlisting}
\begin{tikzpicture}[remember picture, overlay]
  \draw[red, thick, rounded corners]
    ($(a_start)+(0,1.7ex)$) rectangle ($(a_end)+(0,-0.8ex)$);
  \node[draw=red, fill=white, rounded corners, anchor=west] 
    (label) at ($(a_start)+(0.8,-0.8)$) {Could be NULL};
  \draw[->, thick, red] 
    (label.north) -- ($(a_start)+(1.8,-0.15)$);
\end{tikzpicture}
\end{minipage}\\[-4mm]
\begin{tabular}{cc}
Human fix & GPT-4o fix \\
\begin{minipage}{.45\textwidth}
\vspace{1mm}
\begin{lstlisting}[{xleftmargin=1em}]
while (replace) {
    @\SetNode{b_start1}@void *rk = replace->key;@\SetNode{b_end1}@ 

    
    if (@\SetNode{b_start2}@rk@\SetNode{b_end2}@ && table->key_cmp(rk, key) == 0) {
        replace->value = val;
        return CC_OK;
    }
}
\end{lstlisting}
\begin{tikzpicture}[remember picture, overlay]
  \draw[red, thick, rounded corners]
    ($(b_start1)+(0,1.7ex)$) rectangle ($(b_end1)+(0,-0.8ex)$);
  \draw[red, thick, rounded corners]
    ($(b_start2)+(0,1.7ex)$) rectangle ($(b_end2)+(0,-0.8ex)$);
  \node[draw=red, fill=white, rounded corners, anchor=west] 
    (label) at ($(b_start1)+(1.5,-0.45)$) {NULL check added};
  \draw[->, thick, red] 
    (label.west) -- ($(b_start2)+(0.2,0.27)$);
\end{tikzpicture}
\end{minipage}
&
\hspace*{9mm}
\begin{minipage}{.45\textwidth}
\vspace{1mm}
\begin{lstlisting}
 while (replace) {
    @\SetNode{c}@

    if (table->key_cmp(replace->key, key) == 0) {
        replace->value = val;
        return CC_OK;
    }
}
\end{lstlisting}
\begin{tikzpicture}[remember picture, overlay]
  \node[draw=red, fill=white, rounded corners, anchor=west] 
    (label) at ($(c)+(2,-0.1)$) {No changes applied};
\end{tikzpicture}
\end{minipage}
\end{tabular}
\caption{GPT-4o-generated code for fixing a null pointer dereference error.}
\label{fig:commit3}
\end{figure*}

\section{Motivational Examples}
\label{sec:prelim}

The fundamental question we ask is ``Why can't LLMs be effective contributors on sizable projects''; once we know how they fail in our setting -- a variety of tasks from a variety of files in a variety of projects -- we can generalize to larger contexts, and proceed to address or mitigate some of the issues.
We motivate our approach by showing the generated code if we asked LLMs 
to produce the code required for typical maintenance and evolution needs. The four tasks are actual commits (bug fixes and feature enhancements) from our examined projects; the LLMs
were given the affected file as context and the actual commit message as prompt. Due to space constraints, we show the code output from all three LLMs for only two tasks, and one LLM for the other two tasks. The first commit 
allows us to check whether an LLM can complete a straightforward task, as the fix is suggested in the prompt and requires changing a single line of code. The second 
and third 
commits involve fixing use-after-free and null pointer dereference errors, which are among the 2024 CWE Top 25 Most Dangerous Software Weaknesses list~\cite{top25CWE2024}. LLMs' ability to solve these tasks is critical, as these vulnerabilities are common in software projects. The last commit 
is a feature enhancement that requires understanding of the data structure used; this task examines LLM's understanding of the context. For these tasks, we use the commit message as part of the prompt to the LLM and attach the affected file as contextual reference. The prompt we used for fixing a bug was ``\textit{Modify the function in the provided C file such that it fixes the following issue: <\textbf{commit message}> in <\textbf{function name}>}''; for implementing a new feature, we used ``\textit{Implement a function in the provided C file: <\textbf{function name}> such that it <\textbf{commit message}>}''.

\subsection{Fixing an Incorrect Index Bug}
\label{sec:prelim:commit1}

This commit, from \textit{Collections-C}~\cite{srdja}, fixes a function that returns an element at the wrong index.\footnote{\url{https://github.com/srdja/Collections-C/commit/5ba7eed}}
\Cref{fig:commit1} shows the bug, the human fix, and the GPT-4o-generated fix.
The bug is shown on line 6; the human fix correctly changed \ls|deque->buffer[index]| to \ls|deque->buffer[p]|. The commit message is: \textit{``The \ls|deque_remove_at| function was returning an incorrect element in certain situations. The correct element is the one at index \ls|p|''.} However, GPT-4o fails to fix the bug; on the right we 
highlight the three issues with the GPT-4o-generated code. First, the calculation of \ls|p| is modified, which leads to accessing the element at the wrong index again. Second, GPT-4o removes certain variables that are irrelevant to the fix. Third, GPT-4o uses an undeclared variable, \ls|deque->head|, which causes a compilation error, despite the fact that the entire file (prior to the commit) was provided as context, including the declaration of the \ls|deque| struct.

\subsection{Fixing a Use-after-free Error}
\label{sec:prelim:commit2}

This commit, from \textit{jansson}~\cite{akheron} involves fixing a potential use-after-free error when generating an error message.\footnote{\url{https://github.com/akheron/jansson/commit/11813f4}} 
The code, shown in \Cref{fig:commit2}, has a path on which a pointer \ls|key| is used as an argument to \ls|set_error| after it has been freed via \ls|jsonp_free(key)|. The fix involves moving the call to \ls|set_error| before invoking the \ls|jsonp_free| function on it, as shown in the human code (center-left). The commit message used for the prompt was: \textit{``Fix a use after free''}. However, all three LLMs fail to identify this problematic issue. In the GPT-4o-generated code, the variable \ls|key| is still being used after it has been freed. Furthermore, GPT-4o attempts to fix the use-after-free error by decrementing the reference count of \ls|value| at line 3, which reflects a misuse of the function, since \ls|value| is not an allocated object in this context and does not have a reference count. Ministral3 replaces the line \ls|goto error| with \ls|return null| and a call to decrement the reference count of \ls|object| at line 3, which does not change the program as the \ls|error| label performs the same operations; this change does not fix the use-after-free error. Qwen3-Coder makes no changes, and hence does not fix the error.

\begin{figure*}[t]
\centering
\begin{tabular}{cc}
Human & GPT-4o \\
\begin{minipage}{.45\textwidth}
\vspace{1mm}
\begin{lstlisting}[{xleftmargin=1em}]
unsigned int json_object_size(const json_t *json)
{
    json_object_t *object;@\SetNode{a1}@
    if(!json_is_object(json))
        return @\SetNode{a2_start}@-1@\SetNode{a2_end}@; 

    object = json_to_object(json);
    return object->hashtable.size;
}
\end{lstlisting}
\begin{tikzpicture}[remember picture, overlay]
  \node[draw=red, fill=white, rounded corners, anchor=west, align=left] 
    (label) at ($(a1)+(0.35,0.25)$) {Implementation differs\\from the specification};
  \draw[red, thick, rounded corners]
    ($(a2_start)+(-0.1,1.7ex)$) rectangle ($(a2_end)+(0.2,-0.8ex)$);
  \node[draw=red, fill=white, rounded corners, anchor=west] 
    (label1) at ($(a2_end)+(1,-0.1)$) {Returns -1 instead of 0};
  \draw[->, thick, red] 
    (label1.west) -- ($(a2_end)+(0.2,0.05)$);
\end{tikzpicture}
\end{minipage}
&
\hspace*{4mm}
\begin{minipage}{.48\textwidth}
\vspace{1mm}
\begin{lstlisting}
unsigned int json_object_size(const json_t *json )
{
    if (!json_is_object(json))
        return 0;

    json_object_t *object = json_to_object(json);


    return @\SetNode{b_start}@hashtable_size@\SetNode{b_end}@(&object->hashtable);
}
\end{lstlisting}
\begin{tikzpicture}[remember picture, overlay]
  \draw[red, thick, rounded corners]
    ($(b_start)+(-0.1,1.7ex)$) rectangle ($(b_end)+(0.05,-0.8ex)$);
  \node[draw=red, fill=white, rounded corners, anchor=west] 
    (label2) at ($(b_start)+(2,0.52)$) {{\it hashtable\_size()} is undeclared};
  \draw[->, thick, red] 
    (label2.west) -- ($(b_start)+(1,0.28)$);
\end{tikzpicture}
\end{minipage} \\[-4mm]
Ministral3 & Qwen3-Coder \\
\begin{minipage}{.45\textwidth}
\vspace{1mm}
\begin{lstlisting}[{xleftmargin=1em}]
unsigned int json_object_size(const json_t *json)
{
    if(!json_is_object(json))
        return 0;
    json_object_t *object = 
        json_to_object(json);

    return @\SetNode{b_start}@hashtable_size@\SetNode{b_end}@(&object->hashtable);
}
\end{lstlisting}
\begin{tikzpicture}[remember picture, overlay]
  \draw[red, thick, rounded corners]
    ($(b_start)+(-0.1,1.7ex)$) rectangle ($(b_end)+(0.05,-0.8ex)$);
  \node[draw=red, fill=white, rounded corners, anchor=west, align=left] 
    (label2) at ($(b_start)+(3,1)$) {{\it hashtable\_size()} is\\undeclared};
  \draw[->, thick, red] 
    (label2.south) -- ($(b_start)+(2,0.3)$);
\end{tikzpicture}
\end{minipage}
&
\hspace*{4mm}
\begin{minipage}{.48\textwidth}
\vspace{1mm}
\begin{lstlisting}
unsigned int json_object_size(const json_t *json )
{
    if (!json_is_object(json))
        return 0;

    const json_object_t *object = 
        json_to_object(json);


    return object->hashtable.@\SetNode{b_start}@entries@\SetNode{b_end}@;
}
\end{lstlisting}
\begin{tikzpicture}[remember picture, overlay]
  \draw[red, thick, rounded corners]
    ($(b_start)+(0,1.7ex)$) rectangle ($(b_end)+(0,-0.8ex)$);
  \node[draw=red, fill=white, rounded corners, anchor=east, align=left] 
    (label2) at ($(b_start)+(2.7,1.45)$) {{\it hashtable.entries} is\\undeclared};
  \draw[->, thick, red] 
    (label2.south) -- ($(b_start)+(0.6,0.28)$);
\end{tikzpicture}
\end{minipage}

\end{tabular}
\caption{LLM-generated code for implementing a size retrieval function.}
\label{fig:commit4}
\end{figure*}

\subsection{Fixing a Null Pointer Dereference}
\label{sec:prelim:commit3}

This commit, from \textit{Collections-C}~\cite{srdja}, fixes a potential null dereference (which could lead to undefined behavior such as a segmentation fault).\footnote{\url{https://github.com/srdja/Collections-C/commit/2843fe2}}
The buggy code, shown in \Cref{fig:commit3} (top), invokes \ls|strcmp| (\ls|table->key_cmp| uses \ls|strcmp| under the hood) on \ls|replace->key|, which could be \ls|NULL|. 
The human fix consisted of adding a \ls|NULL| check on \ls|replace->key| to prevent the null pointer dereference. The commit message used for the prompt was: \textit{``Fixes a bug that would be triggered by add, get and remove hashtable operations when a key whose hash value would resolve to table index 0 was used while NULL key was already present in the table''}. \Cref{fig:commit3} demonstrates one of the three functions that GPT-4o was asked to fix. 
GPT-4o did not make any changes to the three functions, even though it confidently stated that the generated code has fixed the issue.

\subsection{Implementing a Function for Size Retrieval}
\label{sec:prelim:commit4}

This commit, from \textit{jansson}~\cite{akheron} implements a new function that returns the hash table size in a \ls|struct| called \ls|json_object_t|.\footnote{\url{https://github.com/akheron/jansson/commit/1e00cd5}}
\Cref{fig:commit4} shows the function, which first checks whether the parameter is an object, then casts it to \ls|json_object_t| if the check passes, and returns \ls|object->hashtable.size|. The commit message only mentions the name of the added function. Therefore, for prompt construction, we use the function description, stated in the human commit: \textit{``Returns the number of elements in *object*, or 0 if *object* is not a JSON object''}. The first difference is that the human implementation deviates from the function description by returning \ls|-1| instead of \ls|0|, whereas all three LLMs follow the prompt and return \ls|0|. The second difference is that GPT-4o's and Ministral3's solutions use an undeclared function, \ls|hashtable_size|, whereas Qwen3-Coder's solution uses an undeclared \ls|struct| member. The function and \ls|struct| are not implemented by the LLMs, but rather the LLMs assume that they exist in the codebase. These cause compilation errors, as expected.

\section{Approach}
\label{sec:approach}

\begin{figure}[t]
\centering
\includegraphics[width=\columnwidth]{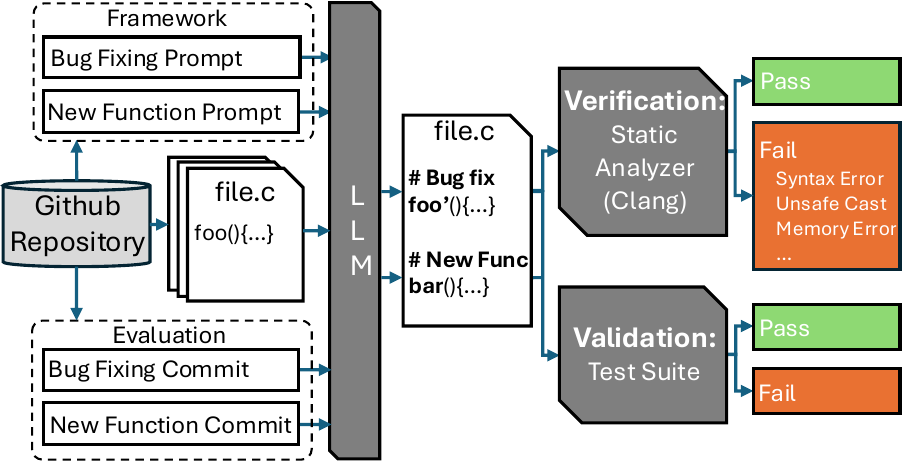}
\caption{Overview of our approach.}
\label{fig:arch} 
\end{figure}

\Cref{fig:arch} shows an overview of our approach: a general {\it framework}, that evaluates a given LLM on a given repository, with a given prompt, and 
an {\it evaluation} that applies this framework to existing commits from our examined projects.

{\it Framework.} The framework is designed to verify and validate LLM-generated code, which can be obtained either (1) by providing a custom prompt or (2) from a commit in a GitHub repository.
The framework takes a source file and a prompt as input which are passed to an LLM to generate a candidate solution. 
This solution is integrated in the original source code and then verified with the Clang static analyzer~\cite{clang} (e.g., for safety issues), followed by validation via the project's test suite (if available).

{\it Evaluation}. We applied our framework to the actual commits from our examined projects. We automated prompt generation by leveraging the actual commit message and the function name affected by the commit. 
Specifically, given a repository and a commit that describes a bug fix or new function implementation, we extract the relevant function (e.g., {\it foo()}, {\it bar()}) from the commit diff and construct a prompt using one of two templates shown in \Cref{sec:prelim}, depending on the nature of the change.
We also identify the changed file (e.g., {\tt file.c}) and retrieve its previous version from the commit's parent, which we then attach as input to the LLM.
The generated function, either a replacement ({\it foo'()}) or a new implementation ({\it bar()}) is inserted into {\tt file.c}. Essentially this yields a code base where the only difference is the LLM-generated patch instead of the original human patch; we run verification and validation on this code base.

{\it No loop.} We deliberately eschewed iterative prompting, i.e., we did not re-prompt in case the LLM-generated code was unsatisfactory.
We made this choice for two reasons. First, LLMs should save developers time, not trade development time for iterative prompting time. Second,
recent research has shown that iterative prompting might degrade the quality of the output or even introduce bugs in bug-free code~\cite{chong2024artificial}.

\subsection{LLMs Used for Evaluation}
\label{sec:approach:llm}

For our study, we used three LLMs: one in the cloud and two local LLMs.
First, we used OpenAI's GPT-4o, 
due to its API availability and widespread adoption.
Second, we used Ministral3 with 14 billion parameters and Qwen3-Coder with 30 billion parameters as local LLMs.
As smaller LLMs are closing the gap with larger LLMs, we wanted to see how well these smaller LLMs perform as compared to larger LLMs such as GPT-4o. In addition, local LLMs are more suitable for the development process, as they incur no token or API fees, work offline, and ensure data privacy.
Using a different LLM involves simply changing the API endpoint. 

\begin{table*}[t]
\centering
\small
\caption{Summary of project, commit, and context properties.}
\resizebox{\textwidth}{!}{\begin{tabular}{lrrrrrrrrrrr}
\toprule
 {\bf Name} & {\bf Stars} & {\bf Forks} & \multicolumn{2}{c}{\bf Commit Timespan} & \multicolumn{2}{c}{\bf Project Size} & \multicolumn{3}{c}{\bf Context (File)  Size} & \multicolumn{2}{c}{\bf \#Commits}\\
            &             &             & \multicolumn{2}{c}{\bf (MM/YY)} & \multicolumn{2}{c}{\bf  (LOC)} & \multicolumn{3}{c}{\bf  (LOC)} & Feature & Bug\\
                 & & &Start &End &Start &End & Min & Median & Max & Enhancements & Fixes\\ \hline
 Collections-C   &  2,900  &     330 & 09/14 & 05/25 &      2,006  &     6,746 &    81 &    493 &     942 & 3  & 17   \\
 packcc          &    371  &      33 & 07/20 & 05/24 &      3,833  &     5,443 & 3,833 &  4,465 &   5,442 & 0  & 5    \\
 libhl           &    442  &     119 & 10/13 & 06/14 &      1,750  &     5,849 &   103 &    557 &     741 & 1  & 19   \\
 jansson         &  3,200  &     833 & 09/09 & 02/15 &      1,882  &     4,142 &    32 &    667 &     779 & 11 & 9    \\
 FFmpeg          & 51,200  &  12,800 & 01/12 & 06/25 &    473,858  & 1,340,782 &    65 &  1,066 &  10,319 & 2  & 73   \\
 wolfSSL         &  2,500  &     876 & 12/23 & 12/24 &  1,345,510  & 1,495,448 &   253 & 23,923 & 125,527 & 0  & 22   \\
 Vsftpd          &   n/a   &  n/a    & 08/09 & 08/21 &     15,302  &    16,415 &    37 &    656 &   2,554 & 4  & 22   \\
 Bison           &   n/a   &  n/a    & 08/18 & 09/21 &     19,681  &    25,480 &    98 &    430 &     824 & 4  & 20   \\
\bottomrule
\end{tabular}
}
\label{tab:project_list}
\end{table*}

\subsection{Projects Used for Evaluation}
\label{sec:approach:projects}

To bolster external validity~\cite{Easterbrook2008}, i.e., choose representative GitHub projects, we proceeded in line with previous research~\cite{herraiz13,jsep13}. Specifically, we selected projects based on four criteria: project size, context size, context size range, and popularity. 
We defined a minimum project size, 1500 LOC, and a median file size of at least 400 LOC
to evaluate LLM performance on realistic setups, and projects likely to come with tests~\cite{kochhar2013empirical}. 
We chose projects where context size range (max/min) was at least 7x, to cover a wide range of file context sizes. 
Finally, we only considered projects above a certain popularity, measured by the number of stars ($>$300) and forks ($>$30), which indicates a diverse contributor list. 

We examined eight open-source projects that met these criteria:
six hosted on GitHub, and two hosted elsewhere.
\Cref{tab:project_list} shows projects' characteristics.
The GitHub projects were:  (1) Collections-C~\cite{srdja} -- a data structure library; (2) packcc~\cite{arithy} -- a parser generator; (3) libhl~\cite{xant} -- a thread-safe data structure library; (4) jansson~\cite{akheron} -- a library to process JSON data; (5) FFmpeg~\cite{ffmpeg} -- a framework for processing audio and video; (6) wolfSSL~\cite{wolfssl} -- a lightweight SSL/TLS library. We also included two open-source projects, the Bison parser generator~\cite{bison} and the Vsftpd FTP server~\cite{vsftpd}; while not hosted on GitHub, their release histories are available, as changelog entries, similar to GitHub commit messages. Moreover, Bison and Vsftpd are the default FTP server and parser, respectively, in popular Linux distributions (Debian/Red Hat and derivatives), which confirms their popularity.
All projects were written mainly in C.

\subsection{Commit Selection Process}
\label{sec:approach:commit}

We selected 212 commits, all targeting the C code in the projects: for GitHub projects we used actual commits, while for Bison and Vsftpd, we selected the tasks from changelog entries (for uniformity, we refer to these entries as commits as well). 
The number of commits, and their kind, for each project, are shown in the last two columns of \Cref{tab:project_list}. 
The commits were selected based on the following criteria.

{\it Single file and single function change.} This criterion keeps the tasks simpler for LLMs.
The rationale behind it is that LLMs are known to struggle on complex tasks, e.g., LLMs are more likely to solve tasks that only take humans 11 minutes or less~\cite{zhang2024cybench}.

{\it Manageable commit size.} The selected commits' median sizes were 4 LOC for bug fixes and 15 LOC for feature enhancement (\Cref{tab:changesInLoC}). When a solution exceeds 12 LOC, 60\% of LLM-generated code tends to be irrelevant to the actual fix~\cite{wang2025towards}; in such cases
chain-of-thought prompting~\cite{wei2022chain} could be used. This prompting technique is recommended by GitHub~\cite{prompt-engineering-copilot} and
has been shown to improve LLMs' ability to perform complex tasks. However, due to the nature of feature enhancement commits -- requiring the implementation of a new function -- their size is slightly higher than for bug fixes.

{\it Well-formed commit messages and changelog entries.} A well-formed commit message contains two main parts: (a) what has changed, and (b) the rationale for the change~\cite{mockus2000identifying, buse2010automatically}. A study shows that typically 56\% 
of open-source projects' commits contain the ``why'' and ``what''~\cite{tian2022what}.

{\it Bug fix variety.} The commits were selected to cover: (1) functional, project-specific issues, and (2) a range of common errors such as null pointer or indexing errors. We believe LLMs should be capable of fixing both common C issues and project-specific bugs.

{\it Feature enhancements.} These involved implementing short, simple, new functions (human equivalent: $\approx$15 LOC) that might call existing functions from the provided context file.

{\it Project lifespan.}
We looked for projects with a long lifespan, to ensure the availability of sufficient commits with a long development history, and capture a wide range of growth phases in the projects' evolution. While project lifespan is omitted from \Cref{tab:project_list} for space reasons, our examined projects' lifespans were at least 5 years (in some cases exceeding 10 years).

{\it Commit dates span pre- and post-cutoff dates.} 
We performed a focused study (GPT-4o only) on the effect of LLM's knowledge cut-off date.
The first period was before GPT-4o's cutoff date (May 2024).
Since these pre-cutoff commits are likely to be in the training data, 
we expect GPT-4o to do well and cite sources.
The second period comes after the training cutoff date, which evaluates GPT-4o's performance on unseen data. 
Ministral3 and Qwen3-Coder do not specify knowledge cutoff dates; the models themselves were released in December 2025 and July 2025, respectively.

\begin{table}[t]
\centering
\small
\caption{LOC changes in the selected commits.}
\begin{tabular}{lcccc}
\toprule
{\bf Commit}            & {\bf Min} & {\bf Max} & {\bf Median}  & {\bf Mean}\\ \hline
Bug Fix                 &  1        & 48        & 4             & 6.61      \\
Feature Enhancement     &  6        & 66        & 15            & 19.96     \\
\bottomrule
\end{tabular}
\label{tab:changesInLoC}
\end{table}

\subsection{Static Analysis}
\label{sec:approach:sa}

For verification 
we use the Clang static analyzer~\cite{clang}, to keep the process scalable for the project size and number of commits we cover. Clang offers a set of default checkers that checks for issues, such as division by zero, dereference of NULL pointers, and use of uninitialized variables. 
We also enable experimental checkers in Clang to check for a wider range of issues, such as out-of-bounds access in string functions and invalid casting between pointers. 

\subsection{Test Suites}
\label{sec:approach:test}

For validation, we use the projects' test suites. Test suites are run to achieve the following two objectives: (1) unit testing, ensuring the LLM-generated code fixes the bug and implements the intended functionality; and (2) regression testing, making sure the LLM-generated code does not break existing functionality.

\subsection{Manual Inspection}
\label{sec:approach:manual}
We also manually inspected the LLM-generated code to confirm it implements the intended feature or fix. This was necessary as we observed that LLM-generated code could pass all the tests even though the correct fix or feature was not implemented, specifically in commits that did not add test cases checking the fix or feature. 
The generated commits were evaluated independently by two PhD students.
We confirmed a 100\% inter-rater agreement rate between the patch assessments.

\begin{table}[t]
\centering
\small
\caption{Categories of issues from static analysis. H: human; G: GPT-4o, M: Ministral3, Q: Qwen3-Coder (LLMs).} 
\begin{tabular}{p{2.7cm}cccc}
\toprule
{\bf Category}              & \multicolumn{4}{c}{\bf \#Issues in Code} \\ 
                            & {\bf H}   & {\bf G}   & {\bf M}   & {\bf Q}       \\ \midrule
Null Dereference            & 23        & 18        & 16        & 9             \\        
Undeclared Identifier       & 0         & 31        & 50        & 20            \\
Unsafe Cast                 & 34        & 64        & 72        & 9             \\
Double Free                 & 0         & 1         & 0         & 2             \\
Use After Free              & 0         & 1         & 1         & 1             \\
Syntax/Semantic             & 0         & 40        & 45        & 1             \\
Uninitialized Variable\!\!  & 4         & 4         & 0         & 0             \\
Empty/Partial               & 0         & 0         & 7         & 44            \\
{\it \!\!\!Total}           & {\it 61}  & {\it 159} & {\it 195} & {\it 97}      \\
\bottomrule
\end{tabular}
\label{tab:SAResultsCategories}
\end{table}

\section{Findings}
\label{sec:findings}

We now discuss the findings of our evaluation, 
structured around four research questions. 
Note that due to computational limitations, we were unable to generate code for 14 out of 22 commits from wolfSSL using Qwen3-Coder when the context file size exceeded 13,000 LOC. Since the context file was included as part of the prompt for local LLMs, a large context file substantially increases the token count, which requires more memory to process the prompt. We used a MacBook Pro with 36GB RAM
for local LLM code generation.

\subsection{RQ1: Can LLMs Generate Syntactically Correct Code?}
\label{sec:rq_compilation_fails}

Failure to compile is one of the prominent issues observed in LLM-generated code.
This is due to the frequent use of undeclared identifiers, such as functions, variables, and \ls|struct| members. Undeclared identifiers account for 31  (19.5\%), 50 (25.6\%), and 20 (20.6\%) issues for GPT-4o, Ministral3, and Qwen3-Coder respectively, as shown in \Cref{tab:SAResultsCategories}. For instance, \Cref{fig:commit4} illustrates the LLMs' assumptions that \ls|hashtable_size| and \ls|hashtable.entries| exist in the codebase, despite their absences in the provided context files.
Another noticeable category is syntax and semantic errors; we observed 40 (25.2\%), 45 (23.1\%), and 1 (1\%) such cases across the three LLMs. 
For example, the use of \ls|case| statement outside \ls|switch| statement, missing curly braces, and treating a variable of type \ls|uint8_t| as a \ls|struct|, are grouped as syntax and semantic issues.
These issues cause compilation errors, as expected (the human-written code does not have such issues, because it would not be accepted as a commit). 
Though such issues are exposed at compile time, fixing them will require a time-consuming human intervention, which undermines the goal of accelerating software development using LLMs.
Note that Qwen3-Coder has an ostensibly lower rate for syntax and semantic errors because it generated empty or partial code for 44 commits, whereas Ministral3 has 7 such cases; uncompilable code vacuously leads to fewer issues during static analysis.

\begin{table}
\centering
\small
\caption{Verification (static analysis) results.}
\resizebox{\columnwidth}{!}{\begin{tabular}{lcccc|C{0.8cm}C{0.8cm}C{0.8cm}}
\toprule
{\bf Projects}              & \multicolumn{4}{c}{\bf \#Issues in Code}                         & \multicolumn{3}{c}{\bf \#Empty/Partial Code}  \\ 
                            & {\bf H}           & {\bf G}       & {\bf M}       & {\bf Q}      & {\bf G}     & {\bf M}   & {\bf Q}    \\ \midrule
packcc                      & 0                 & 0             & 1             & 0            & 0           & 0         & 0          \\        
jansson                     & 0                 & 1             & 17            & 5            & 0           & 0         & 0          \\
Collections-C               & 0                 & 8             & 0             & 3            & 0           & 3         & 3          \\
libhl                       & 3                 & 4             & 22            & 5            & 0           & 0         & 7          \\
FFmpeg                      & 56                & 128           & 99            & 20           & 0           & 0         & 26         \\
wolfSSL                     & 0                 & 5             & 20            & 0            & 0           & 2         & 7          \\ 
Bison                       & 1                 & 12            & 18            & 10           & 0           & 0         & 1          \\
Vsftpd                      & 1                 & 1             & 11            & 10           & 0           & 2         & 0          \\
{\it \!\!\!Total}           & {\it 61}          & {\it 159}     & {\it 188}     & {\it 53}     & {\it 0}     & {\it 7}   & {\it 44}   \\
\bottomrule
\end{tabular}
}
\label{tab:SAResults}
\end{table}

\subsection{RQ2: Why Does Code Fail Static Verification?}

We performed static analysis on both human-written and LLM-generated code for the selected commits. \Cref{tab:SAResults} shows that GPT-4o-generated code has a total of 159 issues (160\% more than human-written code), whereas Ministral3-generated code has a total of 188 issues (208\% more). Qwen3-Coder has a total of 53 issues (13.1\% less than human-written code); however, 44 of its commits are either empty or contain only partial code.
The numerous issues found in LLM-generated code are presented in \Cref{tab:SAResultsCategories} and discussed next.

{\bf NULL Pointer Dereferences.}
Null pointer dereferences are severe issues that can cause denial of service, which is especially a concern if the error happens in low-level systems software. For instance, as shown in \Cref{fig:commit3}, GPT-4o's failure to fix the issue could lead to a program crash (segmentation fault).

\begin{table}[t]
\centering
\caption{Test suites results on LLM-generated code. \# Tests means the total number of tests available.}
\resizebox{\columnwidth}{!}{\begin{tabular}{lccc|ccc|ccc|c}
\toprule
{\bf Projects}      & \multicolumn{3}{c}{\bf Pass}  & \multicolumn{3}{c}{\bf Fail}  & \multicolumn{3}{c}{\bf Compile error} & {\shortstack{\bf \#Tests}}   \\ 
                    & G         & M         & Q     & G     & M     & Q             & G         & M         & Q                 \\ \midrule
packcc              & 2         & 1         & 3     & 1     & 2     & 0             & 2         & 2         & 2                 & 5                     \\        
jansson             & 15        & 12        & 15    & 4     & 4     & 4             & 1         & 4         & 1                 & 20                    \\
Collections-C       & 16        & 15        & 13    & 2     & 2     & 1             & 2         & 3         & 6                 & 20                    \\
libhl               & 13        & 10        & 10    & 6     & 7     & 2             & 0         & 2         & 7                 & 19                    \\
FFmpeg              & 56        & 34        & 27    & 0     & 6     & 3             & 0         & 16        & 26                & 56                    \\
wolfSSL             & 17        & 7         & 2     & 0     & 0     & 0             & 0         & 10        & 15                & 17                    \\ 
Bison               & 5         & 5         & 7     & 6     & 12    & 11            & 9         & 3         & 2                 & 20                    \\
{\it \!\!\!Total}   & {\it 124} & {\it 84}  & {\it 77}  & {\it 19} & {\it 33} & {\it 21} & {\it 14} & {\it 40} & {\it 59} & {\it 157}          \\
\bottomrule
\end{tabular}
}
\label{tab:testSuitesResults}
\end{table}

\begin{table*}[t]
\centering
\small
\caption{Manual inspection results for LLM-generated patches. `Success Rate'=fraction of patches judged correct by the inspection.}
\resizebox{\textwidth}{!}{\begin{tabular}{lp{5.6cm}|ccc|ccc|ccc|ccc|ccc|ccc|ccc|ccc}
\toprule
 &
 & \multicolumn{3}{c|}{\!\!packcc\!\!}
 & \multicolumn{3}{c|}{\!\!jansson\!\!}
 & \multicolumn{3}{c|}{\!\!Collec.-C\!\!}
 & \multicolumn{3}{c|}{\!\!libhl\!\!}
 & \multicolumn{3}{c|}{\!\!FFmpeg\!\!}
 & \multicolumn{3}{c|}{\!\!wolfSSL\!\!}
 & \multicolumn{3}{c|}{\!\!Bison\!\!}
 & \multicolumn{3}{c}{\!\!Vsftpd\!\!} \\
 &
 & \!\!G\!\! & \!\!M\!\! & \!\!Q\!\!
 & \!\!G\!\! & \!\!M\!\! & \!\!Q\!\!
 & \!\!G\!\! & \!\!M\!\! & \!\!Q\!\!
 & \!\!G\!\! & \!\!M\!\! & \!\!Q\!\!
 & \!\!G\!\! & \!\!M\!\! & \!\!Q\!\!
 & \!\!G\!\! & \!\!M\!\! & \!\!Q\!\!
 & \!\!G\!\! & \!\!M\!\! & \!\!Q\!\!
 & \!\!G\!\! & \!\!M\!\! & \!\!Q\!\! \\
\midrule

\!\!\!\multirow{2}{*}{\begin{turn}{90}{\textcolor{green!60!black}{Correct}}\end{turn}}\!\!\!\!
&\!\!\textcolor{green!60!black}{Identical to Human Solution}\!\!
&\!0\!&\!0\!&\!0\!&\!3\!&\!2\!&\!5\!&\!6\!&\!8\!&\!9\!&\!6\!&\!4\!&\!4\!&\!3\!&\!1\!&\!5\!&\!1\!&\!0\!&\!0\!&\!0\!&\!0\!&\!0\!&\!1\!&\!2\!&\!3\!\\ [1.5mm]
&\!\!\textcolor{green!60!black}{Different from Human, Appears Correct}\!\!
&\!2\!&\!0\!&\!2\!&\!9\!&\!4\!&\!3\!&\!6\!&\!2\!&\!1\!&\!5\!&\!2\!&\!3\!&\!7\!&\!3\!&\!4\!&\!2\!&\!0\!&\!0\!&\!10\!&\!4\!&\!7\!&\!8\!&\!2\!&\!2\!\\

\midrule

\!\!\!\multirow{5}{*}{\begin{turn}{90}{\textcolor{red!70!black}{Incorrect}}\end{turn}}\!\!\!\!
&\!\!\textcolor{red!70!black}{Partial Fix}
&\!0\!&\!0\!&\!0\!&\!0\!&\!0\!&\!0\!&\!0\!&\!0\!&\!0\!&\!1\!&\!2\!&\!1\!&\!12\!&\!4\!&\!7\!&\!0\!&\!0\!&\!0\!&\!1\!&\!1\!&\!1\!&\!0\!&\!0\!&\!0\!\\

&\!\!\textcolor{red!70!black}{LLM Did Nothing (Empty Patch)}\!\!
&\!0\!&\!0\!&\!0\!&\!1\!&\!0\!&\!1\!&\!1\!&\!5\!&\!5\!&\!0\!&\!1\!&\!9\!&\!8\!&\!12\!&\!21\!&\!2\!&\!6\!&\!7\!&\!1\!&\!0\!&\!1\!&\!2\!&\!2\!&\!3\!\\

&\!\!\textcolor{red!70!black}{Wrong Solution, Failed to Fix or Implement}\!\!
&\!2\!&\!5\!&\!3\!&\!5\!&\!10\!&\!10\!&\!4\!&\!5\!&\!1\!&\!7\!&\!8\!&\!3\!&\!25\!&\!48\!&\!35\!&\!16\!&\!2\!&\!1\!&\!6\!&\!15\!&\!12\!&\!11\!&\!12\!&\!11\!\\

&\!\!\textcolor{red!70!black}{Deleted Code Unrelated to the Task}\!\!
&\!1\!&\!0\!&\!0\!&\!1\!&\!0\!&\!0\!&\!0\!&\!0\!&\!1\!&\!1\!&\!2\!&\!0\!&\!2\!&\!3\!&\!3\!&\!0\!&\!0\!&\!0\!&\!1\!&\!0\!&\!1\!&\!3\!&\!1\!&\!0\!\\

&\!\!\textcolor{red!70!black}{Uncompilable Code (Undeclared Identifiers)}\!\!
&\!0\!&\!0\!&\!0\!&\!1\!&\!4\!&\!1\!&\!3\!&\!0\!&\!3\!&\!0\!&\!1\!&\!0\!&\!18\!&\!4\!&\!0\!&\!1\!&\!1\!&\!0\!&\!5\!&\!4\!&\!2\!&\!1\!&\!7\!&\!7\!\\

\midrule
\midrule

\multicolumn{2}{l}{\it Success Rate (GPT-4o)}
& \multicolumn{3}{c}{\it 40\%}
& \multicolumn{3}{c}{\it 60\%}
& \multicolumn{3}{c}{\it 60\%}
& \multicolumn{3}{c}{\it 55\%}
& \multicolumn{3}{c}{\it 13.3\%}
& \multicolumn{3}{c}{\it 15\%}
& \multicolumn{3}{c}{\it 41.8\%}
& \multicolumn{3}{c}{\it 34.6\%} \\

\multicolumn{2}{l}{\it Success Rate (Ministral3)}
& \multicolumn{3}{c}{\it 0\%}
& \multicolumn{3}{c}{\it 30\%}
& \multicolumn{3}{c}{\it 50\%}
& \multicolumn{3}{c}{\it 30\%}
& \multicolumn{3}{c}{\it 5.3\%}
& \multicolumn{3}{c}{\it 0\%}
& \multicolumn{3}{c}{\it 16.7\%}
& \multicolumn{3}{c}{\it 15.4\%} \\

\multicolumn{2}{l}{\it Success Rate (Qwen3-Coder)}
& \multicolumn{3}{c}{\it 40\%}
& \multicolumn{3}{c}{\it 40\%}
& \multicolumn{3}{c}{\it 50\%}
& \multicolumn{3}{c}{\it 35\%}
& \multicolumn{3}{c}{\it 12\%}
& \multicolumn{3}{c}{\it 0\%}
& \multicolumn{3}{c}{\it 29.2\%}
& \multicolumn{3}{c}{\it 19.2\%} \\

\bottomrule
\end{tabular}
}
\label{tab:manualInspection}
\end{table*}

{\bf Unsafe Casts.}
This category has the highest proportion for both GPT-4o (40.3\%) and Ministral3 (36.9\%), and accounts for 9.3\% for Qwen3-Coder.
Unsafe casts introduce a range of potential vulnerabilities: data loss, crashes and invalid reads or writes to unintended memory regions. For instance, the LLM-generated code \ls|*dst++ = replacements[i]|, which attempts to assign a \ls|char *| (\ls|replacements[i]|) to a \ls|char| (\ls|dst|) causes an unsafe, pointer-to-character cast.

{\bf Free-Related Vulnerabilities.}
We observed two classes of free-related vulnerabilities: double free and use-after-free. Double free occurs when an allocated region is freed twice, while use-after-free involves accessing a variable after its memory has been deallocated. Both vulnerabilities can lead to undefined behavior.

{\bf Uninitialized Variables.}
Variables that are declared but not assigned to (prior to use) have undefined behavior, with various manifestations across compilers or machines, e.g., as program crashes.

\subsection{RQ3: Validation: Why is Code Incorrect Even though It Passes Verification?}

In addition to the static analysis-based verification, we also performed validation:
we ran the projects' test suites, and followed that with a manual inspection.

\begin{figure*}[t]
\centering
\begin{minipage}{0.975\textwidth}
\centering
Bug
\begin{lstlisting}[xleftmargin=1em]
if (!prev) {
    @\SetNode{a1_start}@ 
    destroy_entry(entry);
    return -1;
}
while (1) {
    if (ATOMIC_READ(prev->next) != ATOMIC_READ(next->node)) {
        @\SetNode{a2_start}@ 
        prev = help_insert(prev, next);
        continue;
    }
\end{lstlisting}
\begin{tikzpicture}[remember picture, overlay]
  \node[draw=red, fill=white, rounded corners, anchor=west] 
    (label1) at ($(a1_start)+(3.4,-0.8)$) {Should release reference here};
  \draw[->, thick, red] 
    (label1.north) -- ($(a1_start)+(3.4,-0.05)$);
  \node[draw=red, fill=white, rounded corners, anchor=west] 
    (label2) at ($(a2_start)+(6.4,-0.6)$) {Should release reference here};
  \draw[->, thick, red] 
    (label2.north) -- ($(a2_start)+(4.9,-0.05)$);
\end{tikzpicture}
\end{minipage}\\[-4mm]
\begin{tabular}{cc}
Human fix & GPT-4o fix \\
\begin{minipage}{.45\textwidth}
\vspace{1mm}
\begin{lstlisting}[{xleftmargin=1em}]
if (!prev) {
    @\SetNode{b_start1}@release_ref(next->refcnt, ATOMIC_READ(next->node));@\SetNode{b_end1}@ 
    destroy_entry(entry);
    return -1;
}
while (1) {
    if (ATOMIC_READ(prev->next) != ATOMIC_READ(next->node)) {
        @\SetNode{b_start2}@release_ref(prev->refcnt, ATOMIC_READ(prev->node));@\SetNode{b_end2}@
        prev = help_insert(prev, next);
        continue;
    }
\end{lstlisting}
\begin{tikzpicture}[remember picture, overlay]
  \draw[red, thick, rounded corners]
    ($(b_start1)+(-0.1,1.7ex)$) rectangle ($(b_end1)+(3.4,-0.8ex)$);
  \draw[red, thick, rounded corners]
    ($(b_start2)+(-0.1,1.7ex)$) rectangle ($(b_end2)+(2.6,-0.8ex)$);
  \node[draw=red, fill=white, rounded corners, anchor=west] 
    (label) at ($(b_start1)+(2.2,-1.2)$) {Reference released};
  \draw[->, thick, red] 
    (label.south) -- ($(b_end2)+(2.0,0.65)$);
  \draw[->, thick, red] 
    (label.north) -- ($(b_end1)+(3.0,-0.15)$);
\end{tikzpicture}
\end{minipage}
&
\hspace*{9mm}
\begin{minipage}{.43\textwidth}
\begin{lstlisting}
if (!prev) {
    @\SetNode{c_start1}@release_ref(next->refcnt, ATOMIC_READ(next->node));@\SetNode{c_end1}@ 
    destroy_entry(entry);
    return -1;
}
while (1) {
    if (ATOMIC_READ(prev->next) != ATOMIC_READ(next->node)) {
        @\SetNode{c_start2}@
        prev = help_insert(prev, next);
        continue;
    }
\end{lstlisting}
\begin{tikzpicture}[remember picture, overlay]
  \draw[red, thick, rounded corners]
    ($(c_start1)+(-0.1,1.7ex)$) rectangle ($(c_end1)+(3.4,-0.8ex)$);
  \node[draw=red, fill=white, rounded corners, anchor=west] 
    (label1) at ($(c_start1)+(2.5,-1.25)$) {Reference released};
  \draw[->, thick, red] 
    (label1.north) -- ($(c_end1)+(2.3,-0.15)$);
  \node[draw=red, fill=white, rounded corners, anchor=west] 
    (label2) at ($(c_start2)+(2.5,-0.65)$) {Missing reference release};
  \draw[->, thick, red] 
    (label2.west) -- ($(c_start2)+(0.5,0.05)$);

\end{tikzpicture}
\end{minipage}
\end{tabular}
\caption{GPT-4o fixed a bug partially.}
\label{fig:llmPartialFix}
\end{figure*}

\begin{figure*}[t]
\centering
\begin{tabular}{cc}
Human fix & GPT-4o fix \\
\begin{minipage}{.45\textwidth}
\vspace{1mm}
\begin{lstlisting}[{xleftmargin=1em}]
case 'r': str[j++] = '\x0d'; break;
case 't': str[j++] = '\x09'; break;
case 'v': str[j++] = '\x0b'; break;
case 'x': 
case 'u':
case '\n': @\SetNode{b_1}@
case '\r': 
default: str[j++] = '\\'; str[j++] = str[i];
\end{lstlisting}
\begin{tikzpicture}[remember picture, overlay]
  \node[draw=red, fill=white, rounded corners, anchor=west] 
    (label1) at ($(b_1)+(0.2,-0.1)$) {Impl. details omitted};
\end{tikzpicture}
\end{minipage}
&
\hspace*{4mm}
\begin{minipage}{.45\textwidth}
\vspace{1mm}
\begin{lstlisting}
case 'r': str[j++] = '\x0d'; break;
case 't': str[j++] = '\x09'; break;
case 'v': str[j++] = '\x0b'; break;
case 'x': @\SetNode{c_1}@ 

@\SetNode{c_2}@

default: str[j++] = '\\'; str[j++] = str[i];
\end{lstlisting}
\begin{tikzpicture}[remember picture, overlay]
  \node[draw=red, fill=white, rounded corners, anchor=west] 
    (label1) at ($(c_1)+(0,0)$) {Impl. details omitted};
  \node[draw=red, fill=white, rounded corners, anchor=west] 
    (label2) at ($(c_2)+(0.2,-0.1)$) {Removed 3 switch cases: \ls|'u'|, \ls|'\n'|, and \ls|'\r'|};
\end{tikzpicture}
\end{minipage}
\end{tabular}
\caption{GPT-4o deleted functional code unrelated to the task.}
\label{fig:llmCodeRemoval}
\end{figure*}

{\bf Test Suites.}
We first ran the test suites on the human-written code for the selected commits; this code passed all test cases, as expected. Next, we replaced the human-written commit with the LLM-generated commit, and re-ran the test suite. We applied this procedure to all projects except Vsftpd, which lacks a test suite. For libhl, we were only able to run 19 test suites out of 20 commits, as one of the libhl versions does not contain a test suite.
Note that we were also unable to run test suites on LLM-generated code that caused compilation errors, such as those resulting from the use of undeclared identifiers (\Cref{sec:rq_compilation_fails}). Therefore, the total number of test suites does not correspond to the total number of selected commits.
In all, for 7 projects, we have a total of 157 test suites available, and we run 98, 117, and 143 test suites (across 98--143 commits) on Qwen3-Coder, Ministral3, and GPT-4o, respectively.

\Cref{tab:testSuitesResults} shows test suite run results. GPT-4o-generated code has the highest number of passing test suites (124), followed by Ministral3 (84), and Qwen3-Coder (77). For failing test suites, Ministral3 has the highest number (33), followed by Qwen3-Coder (21), and GPT-4o (19). 
The reasons for failing test suites include not passing all the test cases and causing the test suite to deadlock and hang.
We observed that {\it a high test suite passing rate does not guarantee code correctness}. Fixes or new features often do not come with new associated test cases in the commits. Therefore, LLMs' failure in fixing bugs or implementing new features may go unnoticed as the relevant test cases are missing from the test suites. 
Note that for Qwen3-Coder, there are 59 commits (37.6\%) that cannot be compiled or used to run the test suites, while there are 40 (25.5\%) such commits for Ministral3 and 14 commits (8.9\%) for GPT-4o.

{\bf Manual Inspection.}
We manually inspected all 212 commits by comparing LLM-generated with human-written code and tracing the LLM-generated code with inputs that trigger the bugs (for bug-fixing commits) and inputs that served as use cases (for feature enhancement commits). \Cref{tab:manualInspection} shows the results. We group incorrect LLM-generated code into 5 categories (category {\it Uncompilable Code due to Undeclared Identifiers} was discussed in \Cref{sec:rq_compilation_fails}).

{\it Partial Fix.} In such cases, the fix requires addressing an issue in, say, $N$ locations in the given function, but the LLM addressed fewer than $N$. For instance, a commit from {\it libhl}~\cite{xant}  fixes a memory leak by releasing references of \ls|next| and \ls|prev| \footnote{\url{https://github.com/xant/libhl/commit/af96d58}}. The bug is shown on lines~2 and~8 in \Cref{fig:llmPartialFix} (top). The human fix involves releasing references of \ls|next| and \ls|prev| on lines~2 and~8, respectively. The commit message is as follows: {\it ``fixed a memory leak triggered by a fault path. the node was overretained''}. However, GPT-4o only released the reference on line~2, i.e., a partial fix.

{\it LLM Did Nothing (Empty Patch).} In such cases, the LLM did not make any changes to the given code, though it was asked to, in the prompt. \Cref{fig:commit3} illustrates this troublesome behavior: GPT-4o outputs code identical to the code prior to the commit, even though GPT-4o is confident in the response that it fixed the bug.

{\it Wrong Solution, Failed to Fix or Implement.} In such cases, the LLM attempted to fix a bug or implement a new feature, but the solution is incorrect; for example, GPT-4o and Ministral3 failed to address an issue in a single location within the function (\Cref{fig:commit2}).  This differs from a partial fix, where the LLM performed one or more fixes, but more changes are needed to fully resolve the issue.

{\it Deleted Code that was Unrelated to the Task.} We also observed that the LLM can delete code unrelated to the task, which in turn breaks existing functionality. For example, a commit from {\it packcc}~\cite{arithy} adds a feature to support escaped backslash.\footnote{\url{https://github.com/arithy/packcc/commit/5211290}} The commit message used is {\it ``Support escaped backslash for both string matching and character class matching''}. However, GPT-4o deleted 3 switch cases -- \ls|u|, \ls|\n|, and \ls|\r| -- which were unrelated to the task, as shown in \Cref{fig:llmCodeRemoval}. This caused the test suites to fail, as expected.

We now discuss the two categories where the LLM-generated code was correct (top rows of \Cref{tab:manualInspection}).

{\it Identical to Human Solution.} For some commits, the LLM generates an exact replica of human-written code. While such behavior can be harmless when the reproduced code is correct, 
we observed that 5 out of 16 citations from GPT-4o across 212 commits cite the wrong resources\footnote{For the commit shown in \Cref{fig:commit3}, the LLM cited 2 GitHub repositories that use versions of {\it Collections-C} prior to the fix, alongside the generated code. These citations were provided as a file that contains code snippets with embedded URLs to their respective GitHub repositories. Anecdotally, this led to the LLM generating an empty patch, as the cited versions do not contain any fixes.} -- using the buggy code prior to the fix -- in the generated code, which left the bugs unfixed. 
This also confirms LLMs' vulnerability to data poisoning (generation of harmful code due to malicious data injected purposely into training data).
For Ministral3 and Qwen3-Coder, the generated code does not contain any citations as they are local LLMs.

{\it Different from Human, Appears Correct.} Here the LLM generates the correct code for the intended fix or feature, though it differs from the human solution. For instance, when checking for integer overflow prior to calling \ls|jsonp_malloc(len + 1)|, GPT-4o's code uses \ls|if (len + 1 == 0)| whereas the human code uses the more robust version \ls|if (len == (size_t) - 1)|.

{\bf Success Rate.}
The last three rows of \Cref{tab:manualInspection} show the overall success rates for each project and LLM.
The lowest success rate, 0\%, is observed for packcc using Ministral3, and wolfSSL using Ministral3 and Qwen3-Coder.
Overall, FFmpeg and wolfSSL have relatively low success rates as these projects have large code bases and large contexts 
(\Cref{tab:project_list}) where LLMs struggle to understand and produce a suitable, or even compilable, patch, as discussed next.

\begin{figure}[t]
\centering
\includegraphics[width=0.92\columnwidth]{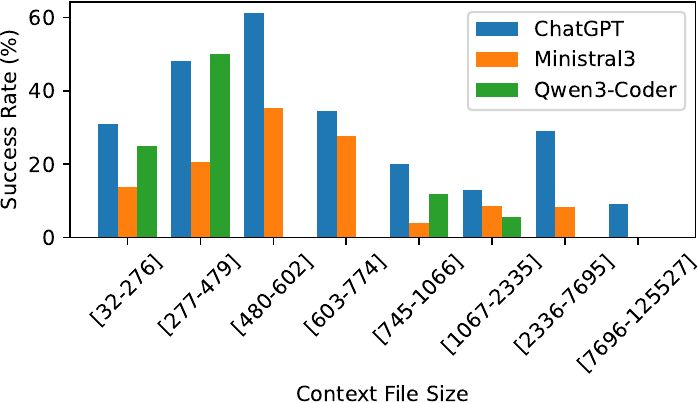}
\caption{Success rate across file sizes (LOC).}
\label{fig:bin-loc} 
\end{figure}

\begin{figure}[t]
\centering
\includegraphics[width=0.92\columnwidth]{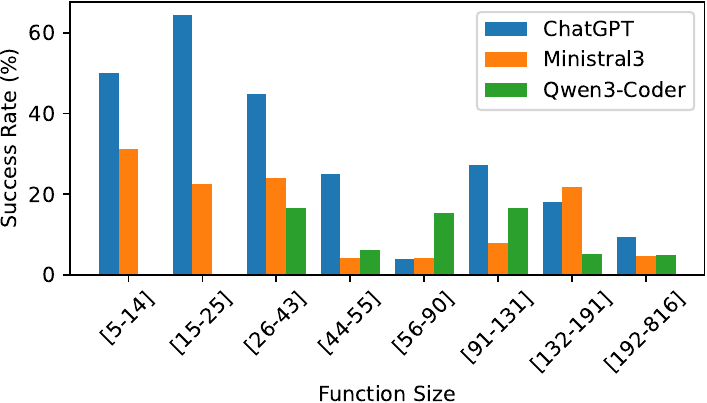}
\caption{Success rate across function size (LOC).}
\label{fig:bin-function} 
\end{figure}

\subsection{RQ4: What Determines the Success and Failure of LLM Code Generation?}
\label{sec:findings:expectedBehavior}

We now discuss key factors contributing to LLM success or failure.

{\it Context File Size and Function Size.} As context file size and function size increase, the success rate first goes up and then drops, as shown in \Cref{fig:bin-loc,fig:bin-function}. We believe the initial increase happens because very small files and functions may not give the LLM enough information to understand the task. With more context, the LLM gets traction -- it can see more of the relevant behavior and dependencies. After a certain point, however, the file or function becomes too large for the LLM to understand reliably. The LLM then has more code to track, the relevant changes become harder to isolate, and the success rate drops.

\begin{table}[t]
\centering
\small
\caption{Statistical analysis: file size and function size.}
\label{tab:t-test}
\resizebox{\columnwidth}{!}{\begin{tabular}{lcccccc}
\toprule
\textbf{} & \multicolumn{3}{c}{\textbf{File Size (LoC)}} & \multicolumn{3}{c}{\textbf{Function Size (LoC)}} \\ \cline{2-7} 
\textbf{Model} & \multicolumn{2}{c}{\textbf{Mean}}& & \multicolumn{2}{c}{\textbf{Mean}} & \\
&  \textbf{Success} \!\!\!\!&\!\!\!\!  \textbf{Failure} \!\!\!\!&\!\!\!\!  \textbf{$p$-value}  \!\!\!\!&\!\!\!\!  \textbf{Success} \!\!\!\!&\!\!\!\!  \textbf{Failure} \!\!\!\!&\!\!\!\! \textbf{$p$-value}  \\
\midrule
GPT-4o \!\!\!\!&\!\!\!\! 1316.6 \!\!\!\!&\!\!\!\! 4972.4 \!\!\!\!&\!\!\!\! 0.0053 \!\!\!\!&\!\!\!\! 50.4 \!\!\!\!&\!\!\!\! 105.7 \!\!\!\!&\!\!\!\! 0.0001\\
Ministral3 \!\!\!\!&\!\!\!\! 690.9 \!\!\!\!&\!\!\!\! 4339.6 \!\!\!\!&\!\!\!\! 0.0003 \!\!\!\!&\!\!\!\! 56.5 \!\!\!\!&\!\!\!\! 94.2 \!\!\!\!&\!\!\!\! 0.0180 \\
Qwen3-Coder \!\!\!\!&\!\!\!\! 551.3 \!\!\!\!&\!\!\!\! 8049.4 \!\!\!\!&\!\!\!\! 0.0005 \!\!\!\!&\!\!\!\! 101.8 \!\!\!\!&\!\!\!\! 151.2 \!\!\!\!&\!\!\!\! 0.0575\\ 
\bottomrule
\end{tabular}
}
\end{table}

We used Welch's t-test~\cite{welch1947generalization} to validate these observed thresholds, by determining if the mean LoC differs significantly between successful and failed generations. As shown in \Cref{tab:t-test}, the results confirm that both file size and function size influence success. GPT-4o exhibited its strongest statistical separation at the function level ($p = 0.0001$):
the mean successful function was 50.4 LoC compared to 105.7 LoC for failed functions, showing that larger function sizes  pose significant challenge for its success. 
Across all models, $p$-values for file sizes range from 0.0003 to 0.0053, with failures often involving files several times larger than successes, e.g., Ministral3's mean failure at 4339.6 LoC vs. 690.1 LoC for success.
For function size, $p$-values are significant (0.0001, 0.018) for GPT-4o as wel as Ministral 3, and indicative for Qwen3-Coder (0.0575). Therefore, the statistical results confirm that LLM's success is heavily dependent on code size.

\begin{table}[t]
\centering
\small
\caption{FFmpeg: success rate before and after the cutoff date.}
\label{tab:cutoff}
\begin{tabular}{llcc}
\toprule
 & & \multicolumn{2}{c}{\textbf{Success Rate}} \\
 & & LOC\textless Median & LOC\textgreater Median \\
\midrule
\multirow{5}{*}{\rotatebox{90}{\shortstack{\textbf{\!Knowledge\!\!} \\\textbf{\!Cutoff Date\!\!}}}} 
 & \multirow{2}{*}{Before}  & 17.4\%        & 13.6\%        \\
 &                          & (4 out of 23) & (3 out of 22) \\[2mm]
 & \multirow{2}{*}{After}   & 14.3\%        & 6.3\%         \\
 &                          & (2 out of 14) & (1 out of 16) \\
\bottomrule
\end{tabular}
\end{table}

{\it Pre- and Post-Knowledge Cutoff Dates.}
We performed a focused study on GPT-4o, which has a significantly larger parameter size than Ministral3 and Qwen3-Coder.
We found that GPT-4o struggles to generate correct solutions 
for commits dated after GPT-4o's knowledge cutoff date (May 2024).
We performed a focused analysis of FFmpeg, which has sufficient post-cutoff commit activity, and show the results in \Cref{tab:cutoff}. 
For commits that precede the knowledge cutoff date the success rate was 
\mbox{13.6--17.4\%} depending on file size, whereas for commits that come after the knowledge cutoff date, the success rate dropped to 6.3--14.3\%.
This finding is concerning, because {\it LLMs will always operate de novo} when writing a patch: they could not have been trained on the very patch being written.

{\it Nature of the Commit.} 
For our second focused study, the LLM, GPT-4o, achieves a lower success rate (29.9\%) for bug-fixing than feature-enhancements commits  (\Cref{tab:featureEnhancementVSBugFix}). Although bug-fixing commits are smaller (median LOC=4), GPT-4o achieves a higher success rate (52\%) on feature-enhancement commits (median LOC=15). 
We believe that locating and fixing a bug is harder for an LLM than implementing a new function because fixing bugs requires understanding the function, in addition to the context file. This aligns with our finding that success rate decreases with larger functions.

\begin{table}[t]
\centering
\caption{Success rate in different commits.}
\small
\begin{tabular}{lcc}
\toprule
                    & {\bf Feature Enhancement}         & {\bf Bug Fix} \\ \midrule
Correct             & 13                                & 56            \\        
Wrong               & 12                                & 131            \\
{\it Success Rate}  & {\it 52\%}                        & {\it 29.9\%}  \\
\bottomrule
\end{tabular}
\label{tab:featureEnhancementVSBugFix}
\end{table}

\section{Discussion}
\label{sec:discussions}

Based on our findings, we provide several recommendations and discuss the limitations of our study.

\subsection{Recommendations}

{\bf Maintain a smaller context file size and function size.} We observed that context file size and function size influence the LLMs' ability to generate correct code for commits. Keeping the context file under 603 LOC (albeit {\it not too small}) and function under 56 LOC increases the likelihood of correct code generation, as shown in \Cref{fig:bin-loc,fig:bin-function}.

{\bf Stricter assessment on unfamiliar-task commits.} The LLM struggles to generate correct code for commits that occur after the LLM's knowledge cutoff date (\Cref{tab:cutoff}). This suggests that for unfamiliar or uncommon software engineering tasks, LLM-generated code should be subject to more rigorous human review and testing.

{\bf Bug-fixing commits require stronger scrutiny.} Bug-fixing commits are more challenging for LLMs, as they require not only locating and fixing the bug, but also understanding both the context file and function. 
Due to their lower success rate (\Cref{tab:featureEnhancementVSBugFix}) we recommend 
stronger scrutiny of LLM-generated 
bug-fixing commits.

{\bf LLM selection for appropriate tasks.} We found that the larger model, GPT-4o, generally performs better than smaller local LLMs, specifically in handling larger context (file/function size). Smaller context and function sizes are often assumed to be beneficial to smaller local LLMs, but in practice they do not reliably lead to correct code generation.
The local LLMs perform well on routine and popular tasks, such as implementing data structures and algorithms, as shown in \Cref{tab:manualInspection}; 
as projects become more specialized, such as in \textit{packcc} and \textit{jansson}, they under-perform. 
Therefore, we suggest a larger model such as GPT-4o for patch generation, Qwen3-Coder (30B parameters in our setting) for exploratory or assistive coding, and Ministral3 (14B parameters) for resource-constrained settings.

\subsection{Limitations}

{\bf Generalizability.} Our study is focused on a single language, C.
This work could be extended to other languages to check for systematic, cross-language issues arising in LLM-based code generation.
Our study examines commits affecting a single file and a single function. The study could be extended to a broader setting, commits affecting multiple files and functions. However, given the already-low success rate in the single file/function setting, we do not expect the success rate to improve when the setting is broader. 

{\bf Commit size.} The selected commits are limited to 66 LOC (\Cref{tab:changesInLoC}) and contain changes to only a single file and a single function. We did not evaluate LLMs' capability in commits that involve changing multiple functions and files. However, our findings reveal that, even in our simplified setting, LLMs' ability to generate correct code deteriorates as the context (file or function) size increases. 

{\bf Availability of commit messages as tasks (prompts).}
Our study relied on the availability of commit messages as task goals, i.e., prompts. When developing {\it de novo}, for bug fixes the prompts can come from bug summaries, after locating/reproducing bugs; for feature improvements, the prompts can come from feature requests.

{\bf Well-formed commit messages.} Our study used well-formed commit messages
(\Cref{sec:approach:commit}). Commits without well-formed messages (e.g., attaching a link to an issue rather than providing a detailed message, or lacking a rationale behind the change) drive the LLM to generate incorrect code. In such cases, the problem is the commit message, not the LLM; interestingly, LLMs can help automate the generation of high-quality commit messages~\cite{wang2021context,liu2022atom,wu2025empirical}.

\section{Related Work}
\label{sec:related}

{\bf LLM Coding Evaluation.}
While Chong et al.~\cite{chong2024artificial} studied the security of LLM-generated code, we study LLMs' capabilities at fixing bugs and implementing new features. They analyzed early changes in Bison and Vsftpd's history; our examined changes are more recent, and do not overlap with the changes and period they studied.
ObfusEval~\cite{zhang2025unseen} uses obfuscated code to test LLMs' capability on unseen tasks; it revealed that the average pass rate of test suites for LLM-generated code can decrease by 15.3\% to 62.5\%, in line with our findings that the LLM struggles to generate code for commits made after its knowledge cutoff date.
Sagodi et al.~\cite{sagodi2024reality} studied automated Java vulnerability repair
using manually engineered prompts.
Other studies examined LLM-generated code's correctness and structural quality~\cite{liu2023,della2025prompt,alves2025quality,coignion2024performance}, robustness~\cite{zhong2024can}, and performance
regressions~\cite{li2024assessing}.
Wang et al.~\cite{wang2025towards} categorized common types of LLM code generation errors using a benchmark dataset.
Huang et al.~\cite{zhang2025repair} introduced ReinFix, an LLM-based program repair framework that guides patch generation using
local and external code bases.
FIXCHECK~\cite{molina2024improving} assesses patch correctness using LLM-assisted test generation, while METAL~\cite{hyun2024metal} 
evaluates consistency of LLM outputs via metamorphic testing.
Orthogonal to these efforts,
we evaluate LLMs' capability in actual, essential software engineering tasks, rather than generic code generation.

{\bf LLM Hallucinations and Non-Determinism.}
LLMs can hallucinate, e.g., by generating syntactically correct but semantically flawed code~\cite{liu2024exploring, zhang2025llm}.
MARIN~\cite{chen2025towards} aims to prevent LLMs from using non-existent APIs and misusing existing ones. Our \Cref{sec:rq_compilation_fails} confirms such hallucinations.
Additionally, the non-deterministic nature of LLMs means that the same prompts can yield different results, causing reproducibility issues~\cite{ouyang2025empirical}.
Turbulence~\cite{honarvar2025turbulence} shows that success on isolated code-generation prompts does not imply robustness, as LLMs frequently fail on closely related prompt instances that differ only in prompt-level constants.

{\bf Prompting with Better Context and Feedback Loop.}
LLM coding task benchmarks often fall short because they do not include the full or correct context (code base).
HumanEvo~\cite{zheng2025humanevo} includes surrounding code and dependencies 
to more accurately assess generated code's quality.
Zhang et al.~\cite{zhang2025instruct} enhance LLMs' ability to modify code snippets by providing richer context and breaking down the tasks.
To address the drawbacks of single-pass code generation, feedback loops have been introduced~\cite{tian2025fixing, chong2024artificial, song2025developing} to perform a cycle of testing and analysis on LLM generated code.
Our framework also considers contextual references, but we do not perform a feedback loop because we evaluate the LLM system as-is.

\section{Conclusions}
\label{sec:conclusions}

Given the rise of LLM-generated code, we study
three LLMs' viability for evolving real, open-source projects, by analyzing actual commits and combining verification\&validation with manual inspection.
LLMs are usable for small files and functions, feature improvements, and tasks similar to prior (seen) ones. In contrast, LLMs
struggle with larger contexts, bug-fixing tasks, and new problems, often producing patches that fail silently. We outline recommendations to alleviate these issues.
Our framework and findings can advance research in both 
enhancing LLMs' code generation capabilities, and verification\&validation approaches for LLM-generated code.

\section*{Acknowledgments}
We thank the reviewers for their valuable feedback. This material is based upon work supported by the National Science Foundation under Grant No. \mbox{CCF-2106710} and an internal NJIT GHRI grant.

\bibliographystyle{unsrtnat}
\bibliography{zephyr,se}

\end{document}